\documentstyle[preprint,aps]{revtex}
\begin{document}

\draft

\preprint{DUKE-TH-95-91;
TMUP-HEL-9510}

\title{Disoriented Chiral Condensate \\
and Strong Electromagnetic Fields}

\author{Hisakazu Minakata}
\address{Department of Physics, Tokyo Metropolitan University \\
Minami-Osawa, Hachioji, Tokyo 192-03, Japan}

\author{Berndt M\"uller}
\address{Department of Physics, Duke University, Durham, NC 27708-0305, USA}

\date{\today}

\maketitle

\begin{abstract}

We discuss the effect of strong electromagnetic fields on chiral
orientation in the framework of the linear $\sigma$ model.  Based on 
lessons we learn from computation of the effective potential at one 
loop, we argue that the chiral U(1) anomaly dominates chiral 
disorientation driven by electromagnetism.  We show that the anomaly 
effect induces a quasi-instantaneous ``kick'' to field configurations 
along the $\pi^0$ direction in relativistic heavy ion collisions.  
We discuss the possibility that a ``kick'', even though small in 
magnitude, can have a substantial effect on the formation of chirally
misaligned domains.
\end{abstract}


There has recently been considerable interest in the possibility of the
formation of a ``chirally misaligned'' domain of space-time as a result
of high energy collisions among protons or heavy nuclei 
\cite{AR91,1,BK92,2,3,4,5}. These domains have been called ``disoriented 
chiral condensates'' (DCC), because they can be formally described as 
localized, coherent excitations of the pion fields corresponding to a local 
rotation of the chiral order parameter of the QCD vacuum. Numerical simulations 
\cite{2,3,4} have shown that such a state can be created spontaneously 
under circumstances where the chiral symmetry of the QCD vacuum has been 
temporarily restored due to a high temperature and then gets broken again 
during the expansion and cooling process.

Because heavy energetic nuclei are sources of strong electric and
magnetic fields, it is of interest to ask whether such fields can have
an effect on the formation of DCC. Some such effects are to be
expected because electromagnetic fields break isospin symmetry, which
is a subgroup of the $SU(2)_L \times SU(2)_R$ group of chiral isospin
transformations. Expressed differently, electromagnetic fields affect
the charged components of the order parameter $\vec\phi=(\sigma,\vec\pi)$ 
of the QCD vacuum.

In this paper we discuss the effect of strong electromagnetic fields on
chiral orientation in the framework of the linear $\sigma$ model.  Based on
lessons we learn from the computation of the effective potential at
one-loop level, we argue that the chiral $U(1)$ anomaly dominates the 
electromagnetic effects on DCC formation.  We will then observe that the 
anomaly effect gives a quasi-instantaneous ``kick'' to the $\sigma$-model 
field configurations along $\pi^0$ direction in relativistic heavy ion 
collisions.  We will attempt to estimate how effective a ``kick'' is in 
changing the chiral orientation of the DCC domain.

As a first step of this investigation we calculate the one-loop
effective potential under the influence of an external static and
uniform electromagnetic field in the framework of the linear
$\sigma$ model.  The Lagrangian density of the model is given by
\begin{equation}
{\cal L} = {1\over 2}\sum \; D_{\mu}\vec\phi D^{\mu}\vec\phi -
{\lambda\over 4} \left(\vert\vec\phi\vert^2 - f_{\pi}^2\right)^2 +
m_{\pi}^2f_{\pi}\sigma
\end{equation}
where $\vec\phi = (\sigma,\vec\pi)$ and
$D_{\mu}=\partial_{\mu}-ieA_{\mu}\hat Q$ denotes the covariant
derivative with charge operator $\hat Q$ of the chiral fields.

We use Schwinger's proper-time formalism \cite{Sch51} to compute the one-loop
effective potential of the the chiral fields.  Assuming that the
charged fields do not develop nonvanishing vacuum values, we integrate 
over the charged field fluctuations (which are the only ones that couple 
to electromagnetic fields) around the charge-neutral $\pi_3$ and $\sigma$ 
background fields.  The result of a one-loop calculation for the effective 
potential can be represented as:
\begin{equation}
{V}_{\rm eff} = {\lambda\over 4} (|\vec\phi|^2 - f_\pi^2)^2 - f_\pi 
m_\pi^2\sigma - {\cal L}'_{\rm em} \label{1}
\end{equation}
where ${\cal L}'_{\rm em}$ summarizes the one-loop effects and takes the form 
\begin{equation}
{\cal L}'_{\rm em} = {\alpha\lambda F^2 \over 24\pi m_\pi^2} 
(\sigma^2 + \pi_3^2 - f_\pi^2) + {7\alpha^2F^4 \over 360m_\pi^4},
\label{2a}
\end{equation}
for weak fields ($eF \ll m_\pi^2$; $F = \sqrt{E^2-H^2}$), and
\begin{equation}
{\cal L}'_{\rm em} = {\lambda|eF| \over 16\pi^2} (\sigma^2 + \pi_3^2 
- f_\pi^2) + {\alpha F^2 \over 4\pi} \log{eF\over m_\pi^2} .
\label{2b}
\end{equation}
for strong fields ($eF \gg m_\pi^2$). 

It is easily seen from these expressions that the electromagnetic fields 
do not alter the orientation of the condensate $\langle\sigma\rangle$ in 
the ground state, a disappointing result.  The reason for this result
is very general; the chiral orientation remains unchanged because the
electromagnetic effect leaves an SO(2) subgroup $(\sigma,\pi_3)$ of
the full SO(4) symmetry of the sigma model invariant.  To show this,
let us write the effective potential in a generic form
\begin{equation}
V_{\rm eff} = U(\sigma^2+\pi_3^2) - f_{\pi} m_{\pi}^2\sigma,
\end{equation}
where $U$ is an arbitrary function.
It is then straightforward to show that the ground state of the
potential is determined as
\begin{equation}
\sigma = {f_{\pi}m_{\pi}^2\over 2U'(\sigma^2)}, \qquad \pi_3 = 0.
\end{equation}
We emphasize that this result is valid not only for static, uniform
electromagnetic fields but also for any space-time dependent fields.
This is due to the fact that the unbroken SO(2) symmetry survives
under any such external electromagnetic fields.

We should note, however, that the SO(2) symmetry is to be broken if
higher-loop corrections of the chiral fields are taken into account.  
We restrict ourselves to the one-loop order, the lowest nontrivial order 
having electromagnetic effects, because we regard the linear $\sigma$ 
model as a low-energy effective theory of QCD.  Therefore, we reach our 
first conclusion that the electromagnetism does not affect the chiral 
orientation of a DCC domain.

In the foregoing discussion we have ignored a fundamental property of
underlying QCD, the existence of the chiral U(1) anomaly (Adler-Bell-Jackiw 
anomaly) \cite{ABJ}.  It gives rise to an interaction term, the Wess-Zumino 
term, of the Nambu-Goldstone bosons with electromagnetic fields in the 
low-energy effective Lagrangian.  The effective potential (\ref{2b}) is 
then modified into
\begin{equation}
{V}_{\rm eff} = U(\sigma^2 + \pi_3^2) - f_\pi m_\pi^2\sigma
- {\alpha\over\pi f_\pi} \vec{E}\cdot\vec{H} \pi_3 , \label{5}
\end{equation}
in which the $SO(2)$ symmetry is broken due to the
Wess-Zumino term.  The condition for the minimum in the variable $\sigma$ 
remains unchanged from (\ref{5}), but the minimum in the $\pi_3$ direction 
now occurs at the position
\begin{equation}
{\pi_3\over\sigma} = {\alpha\over\pi f_\pi^2 m_\pi^2} \vec{E}\cdot\vec{H},
\label{6}
\end{equation}
reflecting the relative strengths of the nonconservation of the axial $U(1)$ 
and the axial $SU(2)$ currents. Inserting the minimum position (\ref{6})
back into the effective potential (\ref{5}) we find
\begin{equation}
{V}_{\rm eff}^{\rm min} = U(\sigma^2 + \pi_3^2) - f_\pi m_\pi^2\sigma
- \left({\alpha \vec{E}\cdot\vec{H} \over \pi f_\pi m_\pi}\right)^2 
{\sigma\over f_\pi}.  \label{7}
\end{equation}
This shows that the energy of the vacuum is always reduced by the anomaly
contribution to the effective potential.

We first discuss what would be the general features of the effects of
the anomaly driven term.  Suppose that a DCC domain is formed during a 
heavy ion collision.  How does the anomaly effect modify the properties of 
the coherent pion emission from DCC?  It must be clear from (\ref{6}) that 
the $\pi^0$ component is enhanced compared with that expected by the isospin
symmetry;  more anti-Centauro events should occur due to the chiral anomaly.

Let us clarify the meaning of this statement.  If the isospin symmetry
is preserved, we have, as the distribution of the neutral-to-all-pion
ratio $R$
\begin{equation}
P(R) = {1\over 2}\; {1\over\sqrt{R}}.
\end{equation}
If we define the Centauro (anti-Centauro) events by $0\le
R\le\varepsilon$ $(1-\varepsilon< R \le 1)$ the ratio of number of
events of anti-Centauro to Centauro is about $\sqrt{\varepsilon}/2$
for small $\varepsilon$.  Our result implies that the anomaly effect
increases this ratio.  To quantify the size of this effect we have to
estimate the magnitude of the anomaly term.

The anomaly term only contributes when parallel electric and magnetic
fields are present. This is, indeed, generally the case in collisions
of heavy nuclei, except in the extreme case of exactly head-on collisions.
Of interest here are collisions at an impact parameter comparable to the
nuclear radii, in which a significant fraction, say one-half, of the 
nucleons of both nuclei participate. Since only long-wavelength
fluctuations of the chiral order parameter can serve as seeds for DCC,
electromagnetic fields created by individual quarks, which dominate
over short distance scales, are irrelevant, and only the coherent fields 
due to the nuclear charges need to be considered.

To get an understanding of the orders of magnitude involved, we can then
estimate the electric and magnetic fields generated by fast moving nuclei 
by using the fields at their surface. In the center-of-mass (c.m.) system 
we have
\begin{equation}
E \approx H \approx {Ze\gamma\over 4\pi R^2} , \label{8}
\end{equation}
where $Z$ and $R$ are the nuclear charge and radius, respectively, and
$\gamma$ is the Lorentz factor in the c.~m.~system. When the nuclei are
not too relativistic, so that they are not Lorentz contracted to less
than the pion Compton wavelength ($R/\gamma > m_\pi^{-1}$), we can
consider the change in the value of the electric and magnetic fields
as quasi-adiabatic. Using (\ref{6}) we find that the change in orientation in
the ($\sigma - \pi_3$) plane is
\begin{equation}
{\pi_3\over\sigma} = \left( {Z\alpha\gamma \over 2\pi f_\pi m_\pi R^2}
\right)^2 \leq  \left( {Z\alpha\over 2\pi f_\pi R} \right)^2
\approx 10^{-3} \label{9}
\end{equation}
for two colliding Pb or Au nuclei.  Therefore, the effect does not
appear to be sizable for nonrelativistic collisions.  

On the other hand, if the nuclei are ultrarelativistic ($E/A \gg 10$ GeV/u), 
they are Lorentz contracted to pancakes far narrower than $m_\pi^{-1}$ and 
the electromagnetic anomaly provide a quasi-instantaneous ``kick'' of the 
ground state away from the $\sigma$-axis. During the brief moment of overlap 
of the nuclei, the effective potential in the $\pi_3$ variable in the 
vicinity of the normal ground state has the form
\begin{equation}
V_{\rm eff}(\pi_3) = {1\over 2} m_\pi^2 \pi_3^2 - \left( 
{Z\alpha\gamma \over 2\pi f_\pi R^2} \right)^2 f_\pi \pi_3 . \label{10}
\end{equation}
The linear term driving the $\pi_3$ field to nonzero values lasts only for
a time of order $R/\gamma$, imparting a kick to the conjugate field momentum
of order
\begin{equation}
\Delta\dot\pi_3 = {(Z\alpha)^2\gamma \over 4\pi^2 f_\pi R^3}\approx 
{1\over 60} m_\pi^2 \label{11}
\end{equation}
for two colliding Au nuclei at RHIC ($\gamma=100$). This results in a
coherent excitation of the chiral order parameter in the $\pi_3$ direction
over a volume of nuclear dimensions. Although the amplitude is quite small
on the scale of $m_\pi$, its coherence over a region much larger than 
$m_\pi^{-1}$ could be quite important because it establishes an
explicit breaking of isospin symmetry in the initial conditions for the
formation of a DCC state. 

One can elaborate these estimations by computing the Lienard-Wiechert
potential of the moving heavy ions under the point-nucleus
approximation.  One obtains
\begin{equation}
\vec E\cdot\vec B = - {2Z^2e^2\over M}\; {\gamma^2 \over
R_1^3R_2^3} (\vec r\cdot \vec L) \label{**}
\end{equation}
where $\vec L=\vec b\times M\vec v$ with ion mass $M$ and impact parameter 
$\vec b$, and
\begin{equation}R_{1,2} = \sqrt{\gamma^2 (z\mp vt)^2+ \left( \vec r_{\perp} 
\pm {\vec b\over 2}\right)^2}.
\end{equation}
The origin of the coordinate system is taken at the collision point.
The expression (\ref{**}) indicates that the anomaly effect vanishes
on the scattering plane and $\pi_3$ is kicked to the opposite
direction in the northern and the southern hemispheres with respect to the
axis defined by $\vec L$.  Using $\vert r \vert\sim \vert b\vert\sim R$ and 
$v\sim 1$ for the sake of an estimate, one obtains essentially the same 
expression as (\ref{10}) except for some enhancement factor of order 10 
in the vicinity of the ions.  

Let us attempt a rough estimate of how the ``kick'' toward the $\pi_3$ 
direction affects DCC formation.  For this purpose we follow
the treatment by Blaizot and Krzywicki \cite{BK94}.  They analyze the 
classical field equations of the linear $\sigma$ model in the $1 + 1$ 
dimensional approximation.  They solve the motion along the proper-time 
$(\tau)$ direction which may be appropriate for ultrarelativistic 
collisions, where initial conditions are expected to be approximately
boost invariant \cite{Bj83}.  In the reasonable limit 
$m_{\sigma}\gg m_{\pi}$ one can ignore the chiral symmetry breaking term.  
In fact, one can verify that coefficients of the quartic term in their 
two-dimensional Lagrangian are larger than $m_{\pi}^2$ by an order of 
magnitude.  One then finds two constant vectors in isospin space (with 
prime denoting the derivative by $\tau$):
\begin{eqnarray}
\vec a &= &\tau (\vec\pi \times \vec\pi') \\
\vec b &= &\tau (\vec\pi\sigma' - \sigma\vec\pi') 
\end{eqnarray}
corresponding to the vector and the axial-vector current conservation,
respectively.

If the anomaly effect is included, the third component of the vector
$\vec b$ is no longer conserved during the encounter of the two nuclei
but obtains an additional contribution
\begin{equation}
\Delta b_3 = - {\alpha\over\pi f_{\pi}^2} \int d\tau\;
\tau(\vec E\cdot \vec B)\sigma.
\end{equation}
Therefore, the effect of the anomaly on $b_3$ vanishes at the chirally
symmetric point $\sigma=0$, and it is maximal at $\sigma=1$.  (Note
that $\sigma$ is rescaled by $f_{\pi}$ in two dimensions.)  By doing
a similar estimate as we did earlier we obtain
\begin{equation}
\Delta b_3 \approx {1\over 8\pi^2} \left( {Z\alpha\over
f_{\pi}R}\right)^2 \approx 10^{-3}
\end{equation} for $\tau \sim R/\gamma$ and $f_{\pi}R\approx 3$.  Thus,
this estimate indicates that the anomaly effect is small at least 
numerically.  

One might conclude that a kick of such tiny magnitude has no chance of 
affecting the DCC formation.  However, we argue that it may well be
effective in triggering the formation of chirally misaligned domain.  
It is natural to suspect that at the pre-DCC stage collision debris
is still hot and the chiral symmetry is restored.  Therefore, the ground
state is at around the top of the Mexican-hat potential.  There are 
thermal fluctuations which trigger the rolling-down motion from the top.
But they trigger the motion in a chirally symmetric way.  There is the
effect of the electromagnetic mass difference between charged and
neutral pions, which may be comparable in magnitude 
--- $(m_{\pi^+}^2-m_{\pi^0}^2)/(m_{\pi^+}^2+m_{\pi^0}^2) \sim 1/30$ --- 
with the effect of the anomaly (\ref{11}). However, the isospin structure
of this term is quite different, $\Delta I=2$ as opposed to $\Delta I=1$
of the anomaly term, and it does not provide a force away from the
axis $\langle\vec\pi\rangle=0$.  We emphasize that the direction of the
``kick'' is uniform over the volume of the collision debris.  Since it
has opposite directions in two hemispheres separated by collision plane
we suspect that the effect may enhance the formation of two DCC domains.
We therefore believe that the anomaly effect, in spite of its smallness,
may play a role in the formation of DCC.

One of the best way of evaluating the anomaly effect on DCC formation 
is to implement its effect into the numerical simulations \cite{2,3,4}.  
The characteristic features of the anomaly ``kick'' into the $\pi_3$ direction, 
such as  ``polarization'' (i.e., the orientation change in sign from 
northern to southern hemispheres), may play an important role in 
incorporating the effect.

To summarize, we have discussed how strong electromagnetic fields
affect chiral orientation within the framework of the linear $\sigma$
model.  We have found that the chiral U(1) anomaly of QCD, represented
by the Wess-Zumino term in the low-energy effective Lagrangian, plays
a crucial role.  We have argued that it produces a ``kick'' to the
field configurations along the $\pi^0$ direction over a volume of
nuclear dimensions.  We then have discussed the possibility that a
small ``kick'' can act as an efficient trigger in the formation of the
DCC domain.  If this is the case the isospin breaking effect may enhance
anti-Centauro over Centauro events in high energy nuclear collisions.

\acknowledgments  This research was supported in part by grants from
the U.S. Department of Energy DE-FG05-90ER40592 and DE-FG02-96ER40945,
the National Science Foundation INT-9315143, and by a Grant-in-Aid for 
Scientific Research for International Scientific Research Program; 
Joint Research No. 07044092 from the Japanese Ministry of Education, 
Science and Culture.  One of us (B.M.) 
gratefully acknowledges the hospitality of the Department of Physics of 
Tokyo Metropolitan University during his visit in July 1994, while the 
other (H.M.) expresses deep gratitude to members of the Physics Department 
of Duke University for their hospitality where this work is completed.

\end{document}